\documentclass{article}
\usepackage{amsfonts}
\usepackage{amssymb}
\usepackage{amsmath}
\usepackage{amsthm}
\usepackage{mathtools}
\begin{document}

\title{Vector modes in Type 3 New GR\\
or\\
\it{!`Do not substitute constraint equations into a Lagrangian!} }

\author{Alexey Golovnev\\
{\small \it Centre for Theoretical Physics, The British University in Egypt,}\\
{\small \it El Sherouk City, Cairo 11837, Egypt.}\\
{\small \it Centre for Cosmology and Science Popularization (CCSP),}\\
{\small \it SGT University, Gurugram, Delhi-NCR, Haryana 122505, India.}\\
{\small  agolovnev@yandex.ru} } 
\date{}

\maketitle

\begin{abstract}

Some time ago, we published the full count of degrees of freedom in the linearised weak gravity limit of arbitrary New GR models. We did it by considering the linear weak gravity equations and presented a thorough analysis with no ambiguity left. A bit later, we generalised it to linear cosmological perturbations and discussed the strong coupling issues that appear already at this level. Recently, there were claims that some dynamical modes had been missed in our work. However, the authors of the new claims did not look at the equations of motion and analysed the quadratic Lagrangian densities instead. In this paper, I take one of the most elementary cases, namely the vector modes in New GR of Type 3, and show what was their mistake that had led them to claiming that those were dynamical.

\end{abstract}

\section{Introduction}

Given a wide variety of modified gravity models on the market, we need to understand their relative properties, starting from the very feasibility of using them for description of anything potentially realistic. In particular, researchers would often look for ghosts in the spectrum of dynamical modes, ready to dismiss a model that features even a single interacting and propagating ghost, i.e. a physical entity of negative kinetic energy. On one hand, I do not agree with the common opinion that ghosts are so deadly for a physical model. I will comment on that in the Conclusions section.

On the other hand, before even posing the questions about the sign of their energy, we must have a clear vision of what are the dynamical modes in a particular theory and whether their number was well-defined, to start with. Strong coupling issues might stand in our way. If such is the case, then the calculations in linearised approximations might be misleading, and the model is actually problematic due to ill-posedness of the Cauchy problem. Still, my aim in this paper is to show that even a linearised investigation must be done accurately, in order to avoid counting some non-existent modes in, those that might appear due to substitutions at the level of the Lagrangian density. I will use a kind of teleparallel gravity for illustrating this point of mine.

New General Relativity is a family of metric teleparallel models given by a (parity-preserving) quadratic Lagrangian in terms of the torsion tensor $T^{\alpha}_{\hphantom{\alpha}\mu\nu}$. I treat it in the pure tetrad approach, that is the spin connection is put to zero, so that the only variable is an orthonormal tetrad $e^a_{\mu}$ taken as a collection of covariantly constant one-forms \cite{meGeom},
\begin{equation}
\label{theconn}
\Gamma^{\alpha}_{\mu\nu}= e^{\alpha}_a \partial_{\mu}e^a_{\nu}, \qquad T^{\alpha}_{\hphantom{\alpha}\mu\nu}=\Gamma^{\alpha}_{\mu\nu} - \Gamma^{\alpha}_{\nu\mu}, \qquad T_{\mu} =  T^{\alpha}_{\hphantom{\alpha}\mu\alpha}.
\end{equation}
The metric is then {\it defined} as
\begin{equation}
\label{themetr}
g_{\mu\nu}=\eta_{ab} e^a_{\mu} e^b_{\nu}, \qquad \eta_{ab}=\mathrm{diag} (+1, -1, -1, -1)
\end{equation}
and is used for raising and lowering the indices. The action functional is
\begin{equation}
\label{themodel}
{\mathcal S} = \int d^4 x\sqrt{-g} \left(\frac{a}{2}\cdot T_{\alpha\mu\nu}T^{\alpha\mu\nu} + b\cdot T_{\alpha\mu\nu}T^{\mu\alpha\nu} - 2c\cdot T_{\mu}T^{\mu}\right)
\end{equation}
with the case of (two times) the Teleparallel Equivalent of General Relativity (TEGR) being $a=b=c=1$.

I will consider vector perturbations around the trivial background of
$$e^a_{\mu}=\delta^a_{\mu}$$
for the Type 3 of these models \cite{NGRm1}, 
\begin{equation}
\label{Type3}
a=b\neq 0\qquad \mathrm{and}\qquad c\neq a.
\end{equation} 
Actually, one also needs to impose that $3c\neq a$, for otherwise it would be Type 7, or Type 8 in a slightly different classification \cite{Shy}, but this is irrelevant for the vector modes.

For the small perturbations, one can define four divergenceless ($\partial_i u_i = \partial_i v_i= \partial_i c_i = \partial_i w_i = 0$) vectors and represent the vectorial sector of perturbed tetrad as
\begin{equation}
\label{tetrpert}
e^0_0  =  1, \qquad e^0_i  =  u_i, \qquad e^i_0 =  v_i, \qquad e^i_j  =  \delta_{ij} + \partial_j c_i+\epsilon_{ijk}w_k.
\end{equation}
Note that the perturbations will be dealt with in the Euclidean manner, so that the position of indices is important in the tetrad and in the spacetime tensors, however the perturbation variables are defined with the lower position of indices only.

In our paper \cite{NGRm1} we have fixed the diffeomorphism gauge of
$$c_i = 0$$
and reparametrised the variables as
\begin{equation}
\label{repar}
{\mathcal M}_i = \frac{u_i -v_i}{2}, \qquad {\mathcal L}_i = \frac{u_i +v_i}{2}, \qquad w_i=\epsilon_{ijk} \partial_j \chi_k
\end{equation}
with new divergenceless ($\partial_i {\mathcal M}_i = \partial_i {\mathcal L}_i= \partial_i \chi_i = 0$) vectors. Since we are discussing perturbation theory with boundary conditions at spatial infinity, as is needed for an unambiguous scalar-vector-tensor decomposition to start with, the extra spatial derivative in the $\chi$-representation of $w$ is harmless.

The linear equations of motion can then be written as \cite{NGRm1}
\begin{equation}
\label{veeqn}
\begin{array}{rcl}
(a+b)\cdot \left({\ddot{\mathcal M}}_i- \bigtriangleup{\mathcal M}_i\right) & = & 0,\\
(a+b)\cdot {\mathcal M}_i - (a-b)\cdot \left({\mathcal L}_i + {\dot\chi}_i  \vphantom{{\ddot{\mathcal M}}_i}\right) & = & 0,\\
(a+b+2c)\cdot {\dot{\mathcal M}}_i - (a+b-2c)\cdot \left({\dot{\mathcal L}}_i + \bigtriangleup \chi_i\right) & = & 0.
\end{array}
\end{equation}
It is obvious that for the Type 3,  $a=b\neq 0$ and $c\neq a$, the equations (\ref{veeqn}) reduce to
\begin{equation}
\label{3eqn}
{\mathcal M}_i =0, \qquad \bigtriangleup \chi_i = - {\dot{\mathcal L}}_i,
\end{equation}
that is one pure gauge vector and two constrained vectors, with no dynamical modes in this sector. One can immediately see the extra gauge freedom that is still there, in the ($\chi,\mathcal L$)-sector, upon having fully fixed the freedom of diffeomorphisms.

Making use of conformal transformations, we have also checked \cite{NGRm2} what happens in cosmological perturbations (around the tetrad $e^a_{\mu} = a(\tau) \cdot \delta^a_{\mu}$ in the conformal time). The gauge symmetry disappears, and the full vector sector becomes constrained. However, no new dynamical mode was seen there. These are the conclusions that have recently been challenged \cite{momo1, momo2}.

In the following Sections, I will introduce the new papers with different claims (Section 2), briefly discuss the issues of choosing a gauge for diffeomorphisms (Section 3), derive the quadratic action in the simple case of weak gravity in vacuum (Section 4), show that the quadratic action which coincides with the action from these works \cite{momo1, momo2} in vacuum does in reality mean no propagating vector modes (Section 5), and then explain how they arrived at different results (Section 6). Finally, in Section 7, I conclude.

\section{The counterclaims}

The paper \cite{momo1} claimed that we had missed some of the propagating modes in our analysis \cite{NGRm1, NGRm2}. Note that the authors of Ref. \cite{momo1} did not even look at the full equations of motion. They analysed the Lagrangian densities and arrived at their results by substituting constraint equations into the Lagrangian. I will explain below that this is the origin of their mistake. I will do it for the vector modes in New GR Type 3, for it is quite easy and illustrative of the problems.

As a potential origin of the mismatch in our results, they mainly mention \cite{momo1} an inappropriate gauge choice and our use of linear order terms only. None of this can be accepted as a proper explanation. Indeed, the linear level of the equations is the same as the quadratic level of the Lagrangian, therefore their second order terms can give us no correction beyond what we have already obtained in our equations. As to the diffeomorphism gauge, in the vector sector, we used the same gauge, $c_i=0$, as they did. Therefore, I will only give brief comments on the gauge choice below.

In order to prove our results \cite{NGRm1} incorrect, it would be necessary to show that there was a mistake in deriving the equations (\ref{3eqn}), so that the real equations did need some Cauchy data in their linear limit. It had not been done. Still, very recently yet another work appeared \cite{momo2}. They give a gauge-invariant support for the claims of Ref. \cite{momo1}, basically attacking the problem in the same way. This is my motivation for giving a pedagogical exposition of what goes wrong in their approach.

\section{On the gauge choices}

The most reliable strategy would be, of course, to not fix any gauge at all and analyse the model in its full. In reality, it would often be too cumbersome. However, in the vector sector of weak gravity, it is actually very simple. Indeed, if we consider the vectorial part of coordinate changes, $x^i \longrightarrow x^i + \zeta_i$, with $\partial_i \zeta_i=0$, then the transformation of the tetrad (\ref{tetrpert}) takes the form of
$$v_i  \longrightarrow  v_i -{\dot \zeta_i}, \qquad c_i  \longrightarrow  c_i -\zeta_i.$$
It is immediately obvious that the gauge-invariant variable is $V_i = v_i - {\dot c_i}$ and a possible gauge choice is $c_i=0$. The nice point is that it is a full gauge fixing, with no remnant freedom, and in this gauge the vector $v_i$ is numerically equal to the gauge-invariant variable $V_i$.

Actually, the same is true of the Newtonian gauge in the scalar sector of linear perturbations, both in weak gravity and for cosmological perturbations. It is a full gauge fixing with the Newtonian potentials in this gauge being numerically equal to the gauge-invariant Bardeen potentials. This is why we always use this gauge for cosmological perturbations, also in teleparallel cosmology \cite{meTo}, in order to have the unambiguous answers. In other words, our gauge choice cannot lead to incorrect results in the scalar sector either.

In the vector sector of weak gravity, it is quite easy to do the full analysis, too. Indeed, since the background torsion tensor vanishes and the Lagrangian is quadratic in it, for the quadratic weak gravity action it is enough to find the torsion tensor up to the first order. For that, we immediately find
\begin{equation}
\label{vecttor}
\begin{array}{rcl}
T_{00i} & = & {\dot u}_i,\\
T_{0ij} & = &  \partial_i u_j - \partial_j u_i,\\
T_{ijk} & = &  \epsilon_{ijl}\partial_k w_l - \epsilon_{ikl}\partial_j w_l,\\
T_{i0j} & = &  \partial_j \left(v_i  -  {\dot c}_i\right) - \epsilon_{ijk}{\dot w}_k.
\end{array}
\end{equation}
I will analyse the action principle in terms of these variables (\ref{vecttor}), and switch to the new parametrisation (\ref{repar}) for explicitly reproducing our equations (\ref{veeqn}).

The only difference between what will be done below and dealing with the full action without gauge fixing is in having $v_i - \dot c_i$ instead of $v_i$ in the Lagrangian. 
Note that the gauge freedom leads to the corresponding two equations not being independent. The $c$-equation is simply a time derivative of the $v$-equation. This is why putting $c_i$ to zero does not make us lose any information. At the same time, the gauge of $v_i=0$ suffers from the remnant gauge freedom of constant $\zeta_i$, while had we put it directly into the action, then even in the physical terms, the new equation would have been more general than the full system really was. This is why one must be cautious about fixing a gauge inside a Lagrangian. One can find pedagogical explanations of it in papers \cite{meHam, meRu, Ru}.

\section{The quadratic perturbed action}

Having found the tensor components (\ref{vecttor}) and fixed the gauge of
$$c_i =0, $$
it is rather straightforward to calculate the quadratic invariants (up to boundary terms):
$$T_{\alpha\mu\nu}T^{\alpha\mu\nu} = -2 {\dot u_i}^2 - 2 u_i \bigtriangleup u_i - 2 v_i \bigtriangleup v_i + 4 {\dot w_i}^2 + 2 w_i \bigtriangleup w_i + 4\epsilon_{ijk} (\partial_i v_j) \dot w_k + \ldots ,$$
$$T_{\alpha\mu\nu}T^{\mu\alpha\nu} = - {\dot u_i}^2 + 2 u_i \bigtriangleup v_i - 2  {\dot w_i}^2 +  w_i \bigtriangleup w_i -2 \epsilon_{ijk} (\partial_i v_j) \dot w_k - 4 \epsilon_{ijk} (\partial_i u_j) \dot w_k + \ldots ,$$
$$T_i= - {\dot u_i} + \epsilon_{ijk} \partial_j w_k, \qquad T_i T^i = -{\dot u_i}^2 +  w_i \bigtriangleup w_i  - 2 \epsilon_{ijk} (\partial_i u_j) \dot w_k + \ldots .$$
I have heavily used many integrations by parts. For example, for the vectorial term, $\epsilon_{ijk} {\dot u_i} \partial_j w_k$ is substituted by $-  \epsilon_{ijk} (\partial_i u_j) \dot w_k$.

Combining all the expressions above in the model of formula (\ref{themodel}) we get the quadratic Lagrangian
\begin{multline}
\label{Lag1}
{\mathcal L} = (2c-a-b)\cdot {\dot u_i}^2 - a\cdot u_i \bigtriangleup u_i - a\cdot v_i \bigtriangleup v_i + 2b\cdot  u_i \bigtriangleup v_i\\
+2(a-b)\cdot {\dot w_i}^2 + (a+b-2c)\cdot w_i \bigtriangleup w_i + 2(a-b)\cdot \epsilon_{ijk} (\partial_i v_j) \dot w_k + 4(c-b)\cdot \epsilon_{ijk} (\partial_i u_j) \dot w_k.
\end{multline}
One can check that, up to integrations by parts, this is precisely the quadratic Lagrangian\footnote{Note that presentation of the Lagrangians is not very reader-friendly in that paper, up to having many terms identically vanishing upon integration by parts, due to vanishing divergences of the vectors.} of the vector sector in the Ref. \cite{momo1} in the limit of vacuum with no universe expansion. The only difference is in renaming the variables and parameters.

Still, contrary to what they claim, the Lagrangian (\ref{Lag1}) does not have any dynamical degrees of freedom in case of Type 3,
$$0\neq a = b \neq c.$$
For convenience, let me write it down explicitly for this case:
\begin{equation}
\label{3Lag1}
{\mathcal L} =2 (c-a)\cdot {\dot u_i}^2 - a\cdot u_i \bigtriangleup u_i - a\cdot v_i \bigtriangleup v_i + 2a\cdot  u_i \bigtriangleup v_i
+ 2(a-c)\cdot w_i \bigtriangleup w_i + 4(c-a)\cdot \epsilon_{ijk} (\partial_i u_j) \dot w_k.
\end{equation}

\section{Still no new dynamical modes}

For now, I will not repeat the mistake of the papers \cite{momo1, momo2}. That is, I will analyse the equations directly, without substituting any of them back into the action. For the general action (\ref{Lag1}) we find
\begin{equation}
\label{Lag1eq}
\begin{array}{rcl}
(a+b-2c)\cdot {\ddot u_i} - a\cdot \bigtriangleup u_i  +  b\cdot \bigtriangleup v_i + 2(c-b)\cdot \epsilon_{ijk} \partial_j {\dot w_k} & = & 0,\\
-a\cdot \bigtriangleup v_i + b\cdot \bigtriangleup u_i + (a-b)\cdot  \epsilon_{ijk} \partial_j {\dot w_k} & = & 0,\\
2(b-a)\cdot {\ddot w_i} + (a+b-2c)\cdot \bigtriangleup w_i + (b-a)\cdot \epsilon_{ijk} \partial_j {\dot v_k} + 2(b-c)\cdot \epsilon_{ijk} \partial_j {\dot u_k} & = & 0
\end{array}
\end{equation}
by varying with respect to $u$, $v$, and $w$ respectively. Note that the whole structure has been obtained in perturbation theory, so that any equation of $\bigtriangleup X =0$ shape will be solved as $X=0$.

One can substitute the change of variables (\ref{repar}) into the equations (\ref{Lag1eq}) in order to reproduce equations (\ref{veeqn}) of our previous work. The second equations of these systems just coincide immediately. Then, if we find $\dot w$ from the second equation (\ref{Lag1eq}) and substitute it into the third equation (\ref{Lag1eq}), we reproduce\footnote{In case of $a=b$, the same result is obtained by solving the second equation as $u_i=v_i$.} the previous third equation (\ref{veeqn}). Combining its time derivative with the first and second equations (\ref{Lag1eq}), one can also arrive at the old first equation (\ref{veeqn}). Therefore, we have proven that the Lagrangian density (\ref{Lag1}) does lead to the same field equations (\ref{veeqn}) as we had before \cite{NGRm1} and does not feature any dynamical modes in the Type 3 case.

To make this statement sound even more reliable, one can start from the particular case of Lagrangian (\ref{3Lag1}) which amounts to putting $a=b$ in the equations (\ref{Lag1eq}). We clearly see that the first and the third equations are no longer independent (recall that, in accordance with the very nature of perturbative treatment with separation into different modes, we assume that all functions decay at spatial infinity), and the whole system reduces to 
\begin{equation}
\label{3Lag1eq}
u_i = v_i, \qquad \bigtriangleup w_i = - \epsilon_{ijk} \partial_j \dot u_k
\end{equation}
that does not have any dynamical mode and coincides with ours (\ref{3eqn}).

Also, it is curious to see that the Type 3 model in the weak gravity limit (\ref{3Lag1})  enjoys a gauge symmetry,
\begin{equation}
\label{extga}
v_i \longrightarrow v_i + \bigtriangleup \xi_i, \qquad u_i \longrightarrow u_i + \bigtriangleup \xi_i, \qquad w_i \longrightarrow w_i - \epsilon_{ijk} \partial_j \dot \xi_k.
\end{equation}
It is an extra gauge freedom of Type 3 which apparently gets broken already upon transition to cosmology \cite{NGRm2} thus showing a kind of strong coupling (in the generalised meaning \cite{sings}) in the Minkowski background.

\section{The way of producing the fake dynamical modes}

The origin of the mistake in the number of modes \cite{momo1, momo2} is in substituting the constraints into the Lagrangian density. Actually, it is not surprising that it happens since the constraints are not purely algebraic. Getting a constraint directly from the variational principle is akin to having a Lagrange multiplier term. Its substitution is similar to using the d'Alembert - Lagrange principle of extremising the action functional in a restricted class of variations. With nonholonomic constraints in classical mechanics, those formulations are generically not equivalent and constitute what is usually called vakonomic and (the standard) nonholonomic mechanics respectively. The field-theory discussion of substitutions into a Lagrangian can be found in an article \cite{Pons}, thanks to an anonymous referee for drawing my attention to it.

Note also that, unlike the remarks in the appendix of (the second version of) the paper \cite{momo2}, my statement on the effects of substitutions is not only about the details of dynamics, but it also concerns the very number of modes. In the same appendix, they admit that there is no propagating vectorial mode in the weak gravity limit, in the sense of absence of spatial derivatives in the equation of their new dynamical mode. However, this global dynamical mode is anyway {\it not} in contradiction with my claims. I consider only the perturbations that decay at spatial infinity (and therefore the decomposition into scalars, vectors and tensors is uniquely defined). Moreover, a global mode is something infinitely less than a proper field theory mode, with only numbers as its Cauchy data instead of arbitrary smooth functions of spatial coordinates. A question about possible global modes is a very interesting one, but it is not the topic of my analysis.

\subsection{The general equations}

For counting the modes, the paper \cite{momo1} employs a series of beautiful transfigurations of the Lagrangians that do change the physical contents of the model. Let me start from their very first step in this direction in the vector sector. They solve the second equation of the system (\ref{Lag1eq}) for $\dot w_i$ and substitute it into the $\epsilon_{ijk} (\partial_i v_j) \dot w_k $ term of the Lagrangian (\ref{Lag1}). The result is a new Lagrangian density
\begin{multline}
\label{newLag1}
{\mathcal L} = (2c-a-b)\cdot {\dot u_i}^2 - a\cdot u_i \bigtriangleup u_i + a\cdot v_i \bigtriangleup v_i \\
+2(a-b)\cdot {\dot w_i}^2 + (a+b-2c)\cdot w_i \bigtriangleup w_i  + 4(c-b)\cdot \epsilon_{ijk} (\partial_i u_j) \dot w_k.
\end{multline}
They celebrate the decoupling of the variable $v_i$. However, this is now an absolutely different model. It leads to a new equation, if $a\neq 0$,
$$v_i=0$$
which does not follow from the initial system of equations (\ref{Lag1eq}) in any way.

In other words, the new dynamical system is absolutely not the same as it used to be. In particular, it requires that ${\mathcal L} = {\mathcal M}$  in our variables (\ref{repar}). At the same time, one can check that it was not the case before the trick. For example, in case of $b=c=0\neq a$ , the equations were $\square{\mathcal M}=\square \chi =0$ and ${\mathcal L} = {\mathcal M} - \dot\chi$, with no need of $\mathcal M$ and $\mathcal L$ being equal to each other.

\subsection{The Type 3 case}

Though the Lagrangian change came from substituting into a term which was not even there for the Type 3, $a=b$, the authors \cite{momo1} continue with the new Lagrangian (\ref{newLag1}),
$${\mathcal L} = 2(c-a)\cdot {\dot u_i}^2 - a\cdot u_i \bigtriangleup u_i + a\cdot v_i \bigtriangleup v_i
 + 2(a-c)\cdot w_i \bigtriangleup w_i  + 4(c-a)\cdot \epsilon_{ijk} (\partial_i u_j) \dot w_k$$
in this case. From the $w$-equation one can find $\partial\dot u$ and use it in the $\epsilon_{ijk} (\partial_i u_j) \dot w_k$ term upon integration by parts, producing a new contribution to the $w_i \bigtriangleup w_i$ term:
$${\mathcal L} = 2(c-a)\cdot {\dot u_i}^2 - a\cdot u_i \bigtriangleup u_i + a\cdot v_i \bigtriangleup v_i
 + 2(c-a)\cdot w_i \bigtriangleup w_i .$$
So, they get two constrained vectors and one dynamical vector, and no gauge freedom (\ref{extga}) which used to be there in the model at hand.

All in all, one can go all the steps from the papers \cite{momo1, momo2} and reproduce the results. The only problem is that those are then about new models with no relation to analysing the New General Relativity itself.

\subsection{A toy model}

Just as a toy model for the effects of lower-derivative substitutions, consider a simple mechanical system of
$$L = \frac12 {\dot x}^2 + xy + \frac12 {\dot z}^2 + x{\dot z} + y{\dot z}.$$
The equations of motion tell us that
$${\dot y} = 0, \qquad {\ddot x} + x= y, \qquad \dot  z = -x.$$

If we substitute ${\dot z} = -x$ for all the occasions of $\dot z$ in the Lagrangian, we get $L = \frac12 \left({\dot x}^2 - x^2 \right)$, so that the resulting system has got no restriction on $y$, while the dynamical variable $x$ has become effectively more restricted,  ${\ddot x} + x=0$, than it used to be. This effect is kind of opposite (more gauge freedom, less Cauchy data) to what we encounted above in the gravity model.

\section{Conclusions}

I have shown that our calculation of the number of degrees of freedom \cite{NGRm1} is fully consistent with the quadratic actions of the paper \cite{momo1}, at least for the Type 3 vector perturbations in the weak gravity limit, and the mismatch in the claims is due to them substituting constraint equations into the Lagrangians.  As a source of confusion, one might note that two of the three authors of the new paper \cite{momo2} have also co-authored another paper \cite{Hell} where a different way of substituting constraints into the Lagrangian had led them to zero Lagrangian in the vector sector of Type 3. Probably, this puzzle alone should make us doubt the whole approach.

Nowadays, we often need to analyse unusual theoretical models of gravity and other fields. I would say that the considerations above entail that we must be very accurate and avoid blindly performing familiar actions without checking what they really mean and do. I discuss some of such issues in Ref. \cite{sings}.

Note also that, in theoretical physics, there is a typical prejudice, that of physical ghosts being absolutely disastrous. These days, it is becoming more and more clear that the question is much more intricate. The best stable example currently available is an exactly solvable mechanical model with two interacting degrees of freedom of opposite signs \cite{Vik1,Vik2}. On one hand, it is not surprising that the exactly solvable model with bounded motion enjoys discrete spectrum of the unbounded Hamiltonian \cite{Vik2} and allows for constructing a "vacuum" state that corresponds to the minimal value of another conserved quantity \cite{Vik1}. On the other hand, it is enough for seeing that there are interesting aspects of the story.

The main argument against asymptotic ghosts in quantum field theory is practically classical, for it is enough to consider a tree-level diagramme of giving birth to two pairs of opposite energy particles. However, despite the infinite phase space volume for the decay and corresponding statistical-physics-type thinking, there are examples of controllable behaviour in ghostful classical field theories, too \cite{VikF}. In the end of the day, the partial differential equations do possess the usual structure, with only the interaction terms suggesting that the equations have come from a variational principle of opposite sign kinetic terms.

All in all, we might need to rethink our attitude towards ghosts. However, there is an even more immediate task before doing that. Namely, we have to learn it accurately, how to find the degrees of freedom, whatever sign their energy be, to start with.

\end{document}